\documentclass[sigplan,screen]{acmart}

\renewcommand\footnotetextcopyrightpermission[1]{}
\setcopyright{acmcopyright}
\usepackage{fancyhdr}
\AtBeginDocument{%
  \providecommand\BibTeX{{%
    \normalfont B\kern-0.5em{\scshape i\kern-0.25em b}\kern-0.8em\TeX}}}

\usepackage{amsmath}
\usepackage{listings}
\usepackage{xcolor}
\usepackage{hyperref}
\usepackage{graphics}
\usepackage{booktabs}
\usepackage{subfig}
\usepackage{float}
\usepackage[font=small, labelfont=bf]{caption}

\graphicspath{{fig/}}

\newcommand{\design}{\texttt{LibASL}}

\begin{document}

\title{Asymmetry-aware Scalable Locking}

\author{Nian Liu$\dagger$$\ddagger$, Jinyu Gu$\dagger$$\ddagger$, Dahai Tang*, Kenli Li*, Binyu Zang$\dagger$$\ddagger$, Haibo Chen$\dagger$$\ddagger$}
\email{{nianliu,gujinyu}@sjtu.edu.cn, {seatang,lkl}@hnu.edu.cn, {byzang,haibochen}@sjtu.edu.cn}
\affiliation{%
  \institution{$\dagger$ Engineering Research Center for Domain-specific Operating Systems, Ministry of Education, China}
  \institution{$\ddagger$ Institute of Parallel and Distributed Systems (IPADS), Shanghai Jiao Tong University}
  \institution{* College of Information Science and Engineering, Hunan University}
  \country{}
}

\renewcommand{\shortauthors}{N. Liu, et al.}

\begin{abstract}
The pursuit of power-efficiency is popularizing asymmetric multicore processors (AMP) such as ARM big.LITTLE, Apple M1 and recent Intel Alder Lake with big and little cores. 
However, we find that existing scalable locks fail to scale on AMP and cause collapses in either throughput or latency, or both,
because their implicit assumption of symmetric cores no longer holds.
To address this issue,
we propose the first asymmetry-aware scalable lock named \design{}.
\design{} provides a new lock ordering guided by applications' latency requirements,
which allows big cores to reorder with little cores for higher throughput under the condition of preserving applications' latency requirements.
Using \design{} only requires linking the applications with it and, if latency-critical, inserting few lines of code to annotate the coarse-grained latency requirement.
We evaluate \design{} in various benchmarks including five popular databases on Apple M1.
Evaluation results show that \design{} can improve the throughput by up to 5 times while precisely preserving the tail latency designated by applications.

\end{abstract}



\maketitle

\section{Introduction}
\label{sec:intro}

Single-ISA asymmetric multicore processor (AMP) combines cores of different computing capacities in one processor~\cite{kumar2003single, kumar2004single} and has been widely used in mobile devices (e.g., ARM big.LITTLE~\cite{biglittle}).
Combining both faster big cores and slower little cores together, AMP is more flexible in accommodating both performance-oriented and energy-efficiency-oriented scenarios,
such as leveraging all cores to achieve peak performance and using little cores only when energy is preferred.
There is also a recent trend to embrace such an architecture in more general CPU processors,
including the desktop and the edge server~\cite{intelbiglittle, applem1, alderlake, amdhybrid}.
As before, applications on AMP need to use locks for acquiring exclusive access to shared data.
However, we observe that existing locks,
including those scalable in the symmetric multicore processor (SMP)~\cite{boyd2012non,mallock} or non-uniform memory access system (NUMA)~\cite{CNA,shuffle,hclh,flatmcs,cohortlock,cst,HBO,hmcs}, 
fail to scale in AMP and cause collapses in either throughput or latency, or both.

\begin{figure}[]
\centering
\subfloat[Throughput collapse.]{
  \includegraphics[width=0.23\textwidth]{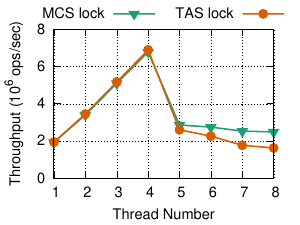}
  \label{fig:thpt_col}
}
\subfloat[Latency collapse.]{
  \includegraphics[width=0.23\textwidth]{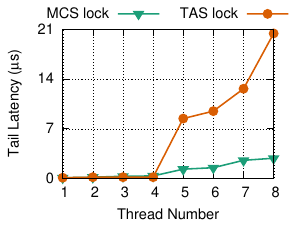}
  \label{fig:lat_col}
}
\vspace{-0.5em}
\caption{Existing locks cause performance collapses on the Apple M1.
M1 has 4 big and 4 little cores.
The first 4 threads are bound to different big cores.
Others are bound to different little cores.
Threads are acquiring the same lock to read-modify-write 4 shared cache lines,
and they will execute a fixed number (400*$2^7$) of \texttt{NOP} instructions between two lock acquisitions.
The throughput is the number of executed critical sections in 1 second.
The latency is the P99 tail latency from acquiring to releasing.
}
\vspace{-1.5em}
\label{fig:ch1_moti}
\end{figure}

After an in-depth analysis, we find the main reason is that those locks (implicitly) assume symmetric cores, which does not hold on AMP.
On the one side, locks that preserve lock acquisition fairness (i.e., give all cores an equal chance to lock), either short-term (e.g., MCS lock~\cite{mcs} passes the lock in a FIFO order) or long-term (e.g., NUMA-aware locks~\cite{CNA,shuffle,hclh,flatmcs,cohortlock,cst,HBO,hmcs} ensure the equal chance in a period),
assume symmetric computing capacity.
Therefore, in AMP, they give the slower little cores the same chance as the big cores to hold the lock,
which introduces the longer execution time of the critical section in little cores to the critical path and causes throughput collapse.
On the other side, locks that do not preserve the acquisition fairness rely on atomic operations to decide the lock holder (e.g., test-and-set spinlock).
They assume a symmetric success rate of the atomic operation when executing simultaneously, which is also asymmetric in AMP.
Thus, those locks are likely to be passed only among one type of core (i.e., either big cores or little cores),
which causes latency collapse even starvation to the others.
Moreover, when the slower little cores have a higher chance to lock,
the throughput also collapses due to the longer execution time of the critical sections on them.
Figure~\ref{fig:ch1_moti} shows the performance collapses on Apple M1.
Both the fair MCS lock and the unfair TAS (test-and-set) lock face throughput collapse when scaling to little cores.
Besides, the TAS lock's latency also collapses and is 6.2x longer than the MCS lock.


Facing the asymmetry in AMP,
it is non-trivial to decide the lock ordering (who can lock firstly) for both high throughput and low latency.
Binding threads only to big cores is an intuitive choice.
However, using little cores can achieve higher throughput under a lower contention.
Besides, it may violate the energy target as the energy-aware scheduler~\cite{eassched} could schedule threads to little cores for saving energy.
Another intuitive approach is to give big cores a fixed higher chance to lock.
However, the throughput and the latency are mutually exclusive in AMP.
It is hard to find a one-size-fits-all (static) setting that can meet the application's latency requirement and improve the throughput simultaneously.

\begin{figure}
\centering
\includegraphics[width=0.45\textwidth]{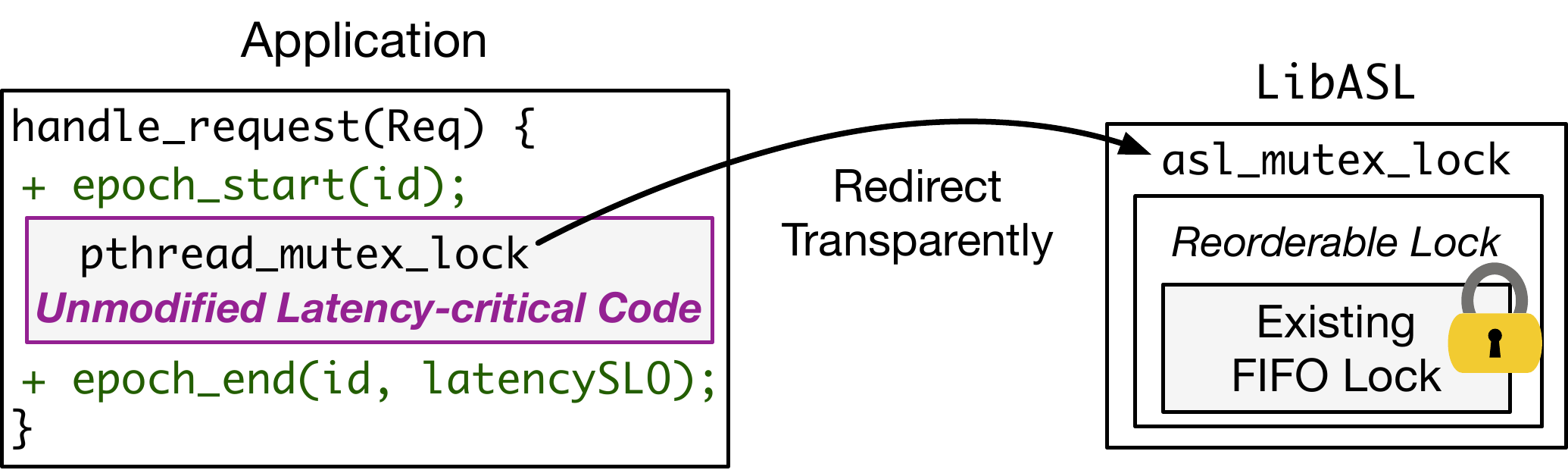}
\vspace{-0.5em}
\caption{\design{} overview.}
\vspace{-1em}
\label{fig:overview}
\end{figure}

In this paper, we propose an asymmetry-aware lock named \design{} as shown in Figure~\ref{fig:overview}.
Rather than ensuring the lock acquisition fairness that causes collapses on AMP,
\design{} provides a new (dynamic) lock ordering guided directly by applications' latency requirements to achieve better throughput under latency constraints.
Atop of a FIFO waiting queue, \design{} allows big cores to reorder (lock before) with little cores as much as possible for higher throughput under the condition that the (reordered) victim will not miss the application's latency target.
To achieve such an ordering, we first design a \textit{reorderable lock},
which exposes the reorder capability as a configurable time window.
Big cores can only reorder with little cores during that time window.
Atop of the \textit{reorderable lock}, \design{} automatically chooses a suitable fine-grained reorder window according to applications' coarse-grained latency requirements through a feedback mechanism.
\design{} provides intuitive interfaces for developers to specify the coarse-grained latency requirements (e.g., request handling procedure) in the form of latency SLO (service level objective, e.g., 99\% request should be complete within 50ms),
which is widely adopted by both academia~\cite{WorkloadCompactor, dvfs,  mittos, AndrewCake} and industry~\cite{googletail, Dynamo, googlecloudSLO}.
To use \design{}, annotating the SLO is the \textbf{only} required effort if latency-critical, which is already clearly defined by applications in most cases.
Non-latency-critical applications can benefit from \design{} without modifications.

We evaluate \design{} in multiple benchmarks including five popular databases on Apple M1,
the only off-the-shelf desktop AMP yet.
Results show that \design{} improves the throughput of pthread\_mutex\_lock by up to 5x (3.8x to MCS lock, 2.5x to TAS spinlock) while precisely maintaining the tail latency even in highly variable workloads.

In summary, this paper makes the following contributions:
\begin{itemize}
\item The first in-depth analysis of the performance collapses of existing locks on AMP.
\item An asymmetry-aware scalable lock \design{}, which provides a new latency-SLO-guided lock ordering to achieve the best throughput the SLO allows on AMP.
\item A thorough evaluation on the real desktop AMP (Apple M1) and real-world applications that confirms the effectiveness of \design{}.

\end{itemize}

\section{Scalable Locking is Non-scalable on AMP}
\label{sec:motiv}

\subsection{Asymmetric Multicore Processor}
\label{sec:amp}

In this section, we introduce the major features of AMP.

First, the asymmetry in AMP is inherent.
Although performance can also be asymmetric in SMP when using DVFS (dynamic voltage and frequency scaling), they can boost the frequency of the lagging core~\cite{TURBO, dvfs-lock-acc, dvfs-identify-cs} while AMP cannot.

Second, asymmetric cores in recent AMP~\cite{inteldeepdive, dynamiqtr, intelbiglittle} are placed in one single cluster and share the same Last Level Cache (LLC).
Thus, communication among cores in AMP is similar to SMP rather than NUMA.

Third, the scheduler (e.g., the energy-aware scheduler in Linux~\cite{eassched}) can place different threads across asymmetric cores~\cite{biglittlecoreswitch, jeff2013big}.
Multi-threaded applications achieve better performance by leveraging all asymmetric cores~\cite{lwnsched} and need to use lock for synchronization among them as before.

\subsection{A Study of Existing Scalable Locks in AMP}
\label{sec:analyze}

Scalable locking in SMP and NUMA has been extensively studied.
However, existing scalable locks are non-scalable on AMP and encounter performance collapses.
The main reason is that existing locks (implicitly) assume symmetric cores, which does not hold in AMP.
There are two major differences between AMP and SMP that cause the collapses.

\emph{First, the computing capacity is asymmetric.}
Little cores spend a longer time executing the same critical section.
Thus, when the lock preserves the acquisition fairness,
it gives little cores an equal chance to hold the lock,
which introduces the longer execution time of critical sections on them to the critical path and causes a throughput collapse. 

\begin{figure}[]
\centering	
\includegraphics[width=0.46\textwidth]{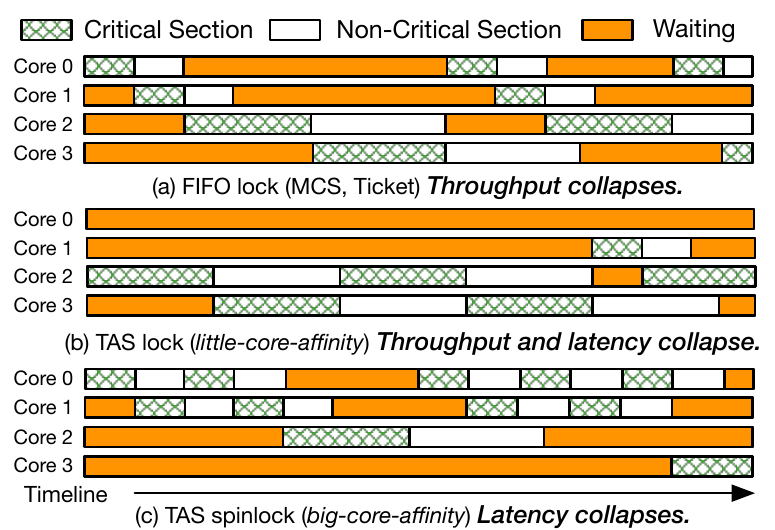}
\caption{An example timeline (from left to right). Core 0/1 are big cores; 2/3 are little cores. More critical sections are executed means higher throughput; longer waiting time leads to longer latency.}
\vspace{-1em}
\label{fig:timeline}
\end{figure}

We explain this problem through an example in Figure~\ref{fig:timeline}(a).
The system includes two big cores (core 0/1) and two little cores (core 2/3), and they are competing for the same lock intensively.
We divide the execution into three parts: executing the critical section, the non-critical section, and waiting for the lock.
As shown in Figure~\ref{fig:timeline}(a),
when ensuring the short-term (i.e., FIFO) lock acquisition fairness (e.g., the MCS lock~\cite{mcs}),
threads hand over the lock in a FIFO order.
As a result, the longer execution time of the critical section in core 2/3 will dominate the critical path and hurt the throughput.

Besides the \textit{short-term} acquisition fairness,
previous work provides \textit{long-term} acquisition fairness to improve the throughput in many-core processors or NUMA while keeping a relatively low latency,
which also hurts the throughput in AMP.
Malthusian lock~\cite{mallock} reduces the contention in the many-core processor for better throughput by blocking all competitors in the waiting queue except the head and the tail.
It achieves long-term fairness by periodically shifting threads between blocking and acquiring.
NUMA-aware locks~\cite{cohortlock, cst, CNA, shuffle,flatmcs, HBO, hclh, hmcs}
batch the competitors from the same NUMA node to reduce the cross-node memory references.
They achieve long-term fairness by periodically allowing different nodes to lock.
All of these locks cannot scale in AMP.
When splitting the asymmetric cores in AMP onto two different nodes,
the long-term fairness will give the \textit{little core nodes} an \textbf{equal} chance to lock as the \textit{big core nodes}.
Thus, similar to MCS, the longer execution time of critical sections in little cores dominate the critical path and causes throughput collapses.

\smallskip

\noindent\textbf{\underline{Implication 1:}}
\emph{
Lock ordering that respects acquisition fairness, either short-term or long-term, is no longer suitable in AMP.
}
In SMP or NUMA, preserving acquisition fairness can prevent starvation and achieve relatively low latency without degrading the throughput but causes collapses in AMP.
Thus, a new lock ordering should be proposed for AMP to meet the latency goal while bringing higher throughput.

\smallskip

\emph{Second, the success rate of atomic operations (e.g., test-and-set, TAS) is asymmetric.}
On some AMP systems (e.g., ARM Kirin970~\cite{hikey970} and Intel L16G7~\cite{intel-l16g7}),
we observe that big cores have a stable advantage over little cores in winning the atomic TAS.
While on other platforms (e.g., Apple M1), the advantage shifts between asymmetric cores\footnote{On Apple M1, when executing TAS back-to-back (higher contention), little cores show a stable advantage.
With the distance between two TAS increased (lower contention), big cores show a stable advantage.}.
Thus, locks that do not preserve acquisition fairness and rely on the atomic operation to decide the lock holder (e.g., TAS lock)
also have the scalability issue in AMP.
Such locks are likely to be held only by one type of core (i.e., either big cores or little cores), which causes a latency collapse even starvation to the others.
Moreover, when the little cores have a bigger chance to hold the lock,
the throughput also collapses due to the longer execution time of the critical sections on them.

As shown in Figure~\ref{fig:timeline}(b),
when little cores show an advantage in winning the atomic TAS,
they have more chance to lock
(we name it as \textit{little-core-affinity}).
Thus, big cores can barely lock.
Besides, the longer execution time of the critical sections on little cores will dominate the critical path and hurt the throughput.
Similarly, when big cores show an advantage (\textit{big-core-affinity}, Figure~\ref{fig:timeline}(c)), little cores will starve.
Nevertheless, it allows big cores to lock before (reorder with) earlier little cores.
Thus more critical sections are executed on faster big cores, which brings higher throughput.

\begin{figure}[]
\centering
\subfloat[Throughput]{
  \includegraphics[width=0.21\textwidth]{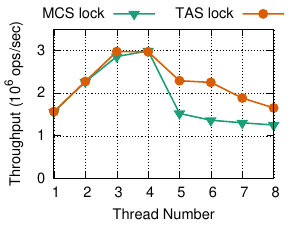}
  }
\subfloat[Latency]{
  \includegraphics[width=0.21\textwidth]{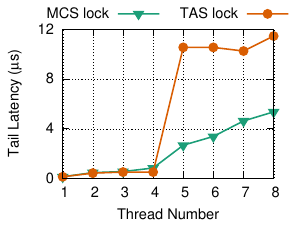}
}
\vspace{-0.5em}
\caption{When TAS lock shows big-core-affinity, it can achieve higher throughput but the latency still collapse.}
\label{fig:ch2_moti}
\end{figure}

\begin{figure}[]
\centering
\vspace{-1em}
\includegraphics[width=0.30\textwidth]{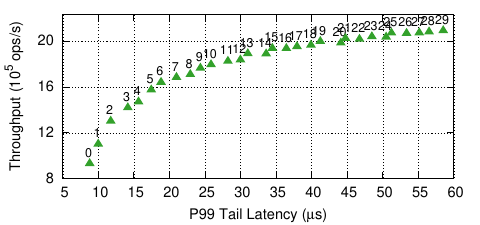}
\vspace{-0.5em}
\caption{Performance when setting different proportions (label on each point). N means big cores have N times higher chance to lock.
The implementation of such an approach is explained in Section~\ref{sec:eval}.
}
\vspace{-1em}
\label{fig:ch2_proportion}
\end{figure}

We validate our observations on Apple M1.
In Figure~\ref{fig:ch1_moti}, we present the case when the TAS lock shows little-core-affinity.
In Figure~\ref{fig:ch2_moti}, we present another scenario when the TAS lock shows big-core-affinity\footnote{
Specifically, read-modify-write 64 cache lines (4 cache lines in Figure~\ref{fig:ch1_moti}) in the critical section.
We identify the affinity by comparing the number of executed critical sections in different types of cores.}.
In both scenarios, the fair MCS lock faces throughput collapses (over 50\% degradation from 4 big cores to all cores), while the unfair TAS lock faces latency collapses.
When the TAS lock shows little-core-affinity in Figure~\ref{fig:ch1_moti},
its throughput also collapses and is 35\% worse than the MCS lock when using all the cores.
However, when the TAS lock shows big-core-affinity in Figure~\ref{fig:ch2_moti},
more critical sections will be executed on the faster big cores, 
bringing 32\% higher throughput than the MCS lock.
The latency of the MCS lock also increases when scaling to little cores due to the longer execution time of critical sections on little cores.
However, it is much shorter than the TAS lock.
These observations still hold in real-world applications.
When the TAS lock shows little-core-affinity in SQLite (detailed in Section~\ref{sec:appbench}),
it has 49\% worse throughput and 1.8x longer tail latency than the MCS lock.
However, when the TAS lock shows big-core-affinity in UpscaleDB,
it has 90\% better throughput yet 2.5x longer tail latency than the MCS lock.


\smallskip
\noindent\textbf{\underline{Implication 2:}}
\emph{
Reordering to prioritize faster cores is indispensable in AMP for higher throughput,
but it must be bounded.
}
When the TAS lock shows big-core-affinity, it reorders big cores with little cores unlimitedly and achieves higher throughput.
However, the unbounded reordering causes a latency collapse.
Thus, the reordering must be bounded for preserving applications' latency requirements.

\begin{figure}[]
\centering
\includegraphics[width=0.4\textwidth]{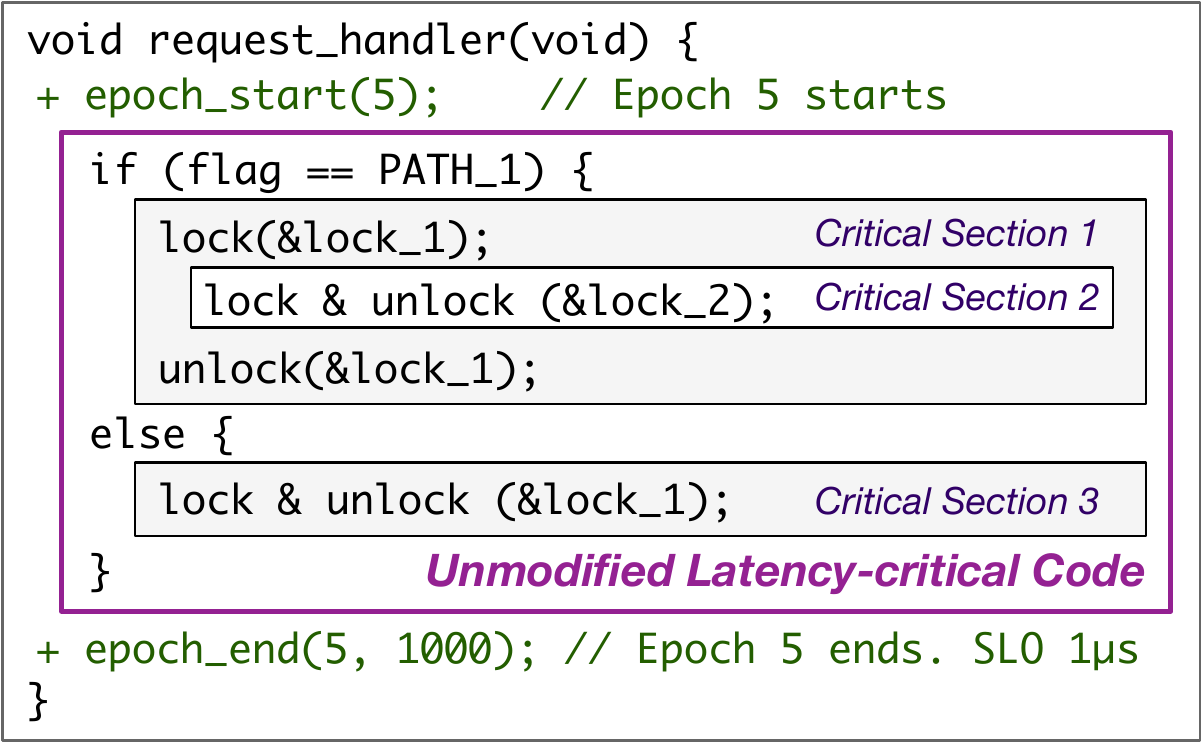}
\vspace{-0.5em}
\caption{Example of using \design{}.}
\label{fig:ch3_usage}
\vspace{-1em}
\end{figure}

\subsection{Strawman Solutions}
\label{sec:strawman}

A straightforward solution is only using big cores.
However, little cores can help achieve higher throughput under a lower contention (e.g., 68\% in Figure~\ref{fig:ch4_exp3}).
Finding the optimal number of cores to run applications is a long-existing problem~\cite{litl, mallock}.
Directly blocking little cores under high contention will cause a latency collapse (Implication 2).
Moreover, since the energy-aware scheduler may schedule threads to little cores for energy purposes,
binding or migrating threads to big cores will violate the energy target.

Another intuitive solution is adopting proportional execution that gives big cores a fixed higher chance to lock.
Figure~\ref{fig:ch2_proportion} shows the performance when setting different proportions.
The throughput and the latency are mutually exclusive in AMP.
A larger proportion brings higher throughput but longer latency.
However, it is unclear whether a specific application prefers throughput over latency (and the extent) or the opposite.
Moreover, since applications' loads may change over time,
the latency is unstable when setting a fixed proportion (not a problem in SMP since the fairness always ensures the lowest latency).
Therefore, it is almost impossible to find a suitable proportion to meet various applications' needs.

\section{Design of \design{}}
\label{sec:design}

\subsection{Overview}

To address the lock scalability problem on AMP,
we propose an asymmetry-aware scalable lock \design{}.
Rather than preserving the lock acquisition fairness,
\design{} provides a new lock ordering guided directly by the applications' latency requirements (i.e., SLO) for better throughput (according to Implication 1).
Atop of a FIFO waiting queue, \design{} allows reordering under the condition that the victim (reordered) will not miss the application's latency SLO.
Thus, big cores can reorder as much as possible with little cores to achieve higher throughput, while little cores can barely meet their latency SLO (according to Implication 2).

To achieve the SLO-guided ordering, bounded reordering is needed.
Thus, we first design a \textit{reorderable lock}, which exposes the bounded reorder capability as a configurable reorder time window atop of a FIFO waiting queue.
Only during the time window,
big cores can reorder (lock before) with little cores.
Once the window expires, no reorder will happen (bounded).
However, it is non-trivial to set a suitable fine-grained window for each lock acquisition
based on the application's coarse-grained latency requirement.
To this end, by proposing a feedback mechanism,
\design{} automatically chooses a suitable reorder window on each lock acquisition according to the coarse-grained latency SLO.

\begin{figure}
\vspace{-1em}
\centering
\includegraphics[width=0.49\textwidth]{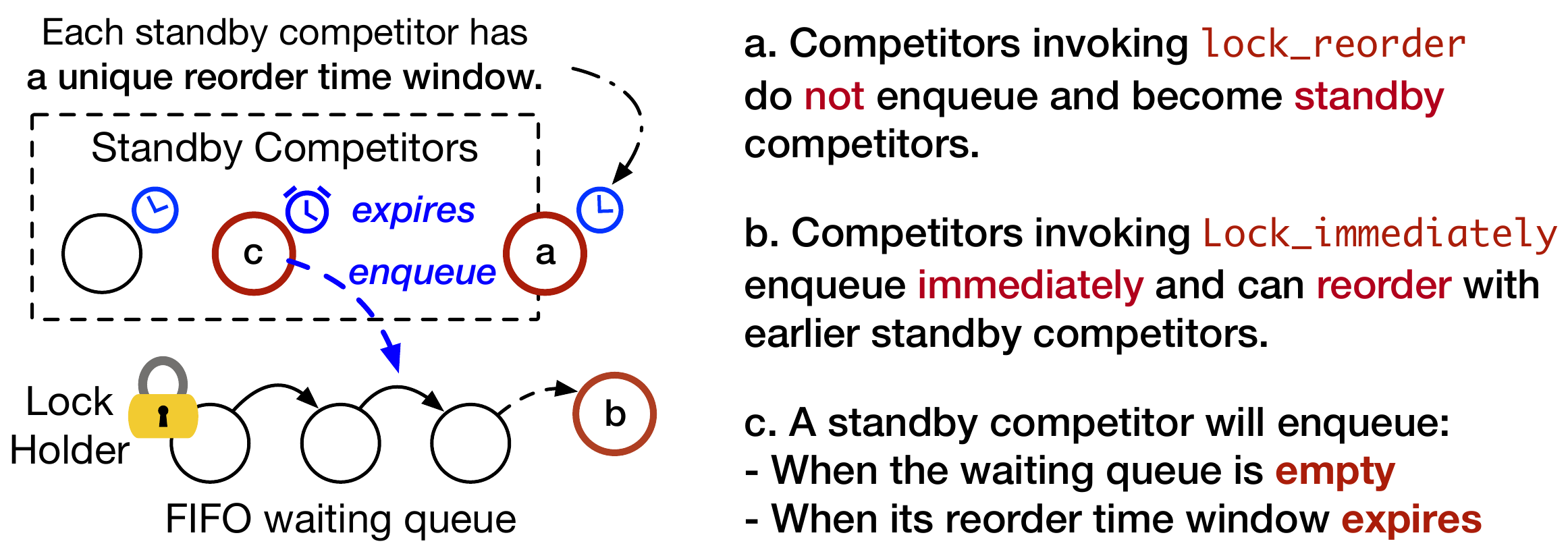}
\label{fig:ch3_lock_interface}
\vspace{-1em}
\caption{Reorderable lock blocks standby competitors and allows other competitors to reorder with them.
}
\vspace{-1em}
\label{fig:ch3_design}
\end{figure}

\smallskip
\noindent\textbf{Usage model.}
\design{} is easy-to-use.
As shown in Figure~\ref{fig:ch3_usage},
\design{} provides \textbf{two intuitive interfaces} to annotate the latency SLO of a certain code block (named as an \textit{epoch}), including \texttt{epoch\_start} and \texttt{epoch\_end}.
Each epoch has a unique epoch id, which is statically given by programmers and should be passed as an argument (e.g., 5 in Figure~\ref{fig:ch3_usage}).
\texttt{epoch\_end} takes another argument, which specifies the latency SLO of the epoch in nanoseconds (e.g., 1000 means the epoch's latency SLO is 1$\mu$s in Figure~\ref{fig:ch3_usage}).
\design{} does not restrict either the number of locks or how the lock is used in an epoch.
Thus, programmers can directly mark a coarse-grained latency SLO without dividing (e.g., the whole request handler in Figure~\ref{fig:ch3_usage})
and create as many epochs as needed.
\design{} leverages weak-symbol replacement to redirect the invocations of pthread\_mutex\_lock transparently.
Therefore, no other modification is required.

\design{} can be used in various applications to improve the scalability under AMP with minimal effort.

\begin{itemize}
	\item For latency-critical applications, annotating their existing coarse-grained SLOs is the \textbf{only} required effort to use \design{}. No other knowledge is required. Such SLOs are defined according to actual latency targets and are commonly available in practice. For example, an interactive app has an SLO of 16.6ms to satisfy the 60Hz frame-rate requirement.
	\item For applications without clear SLOs, \design{} provides a profiling tool that generates a latency-throughput graph (e.g., Figure~\ref{fig:ch4_exp2}) to help choose suitable SLOs. The graph is generated by automatically iterating different SLO settings inside a given SLO range.
	\item Non-latency-critical applications can transparently use \design{} with no SLO. \design{} directly uses a default (fairly loose) reorder time window to maximize the throughput without causing starvation.
\end{itemize}

\subsection{Reorderable Lock}

The reorderable lock exposes the bounded reorder capability atop of existing FIFO locks (e.g., the MCS lock).
It provides two interfaces to acquire the lock, including \texttt{lock\_reorder} and \texttt{lock\_immediately}.
Figure \ref{fig:ch3_design} presents the behavior of both interfaces.
Competitors invoking \texttt{lock\_immediately} will be appended to the tail of the waiting queue immediately.
Competitors invoking \texttt{lock\_reorder} will be regarded as standby competitors.
If the waiting queue is empty, standby competitors can enqueue and then become the lock holder.
Otherwise, the standby competitors are blocked.
Each standby competitor has a unique reorder time window,
which is an argument of \texttt{lock\_reorder}.
Other competitors can reorder with them and lock earlier during that window.
Thus, the reordering is bounded by that window.
A standby competitor will enqueue once its reorder window expires.

Algorithm~\ref{alg:reorder_impl} shows the implementation of the reorderable lock.
When calling \texttt{lock\_immediately}, the competitor will directly enqueue (line 2) by using the lock interface (\texttt{lock\_fifo}) of the underneath replaceable FIFO lock (e.g., MCS).
\texttt{lock\_reorder} takes an argument \texttt{window} that specifies the length of the reorder window in nanoseconds.
When calling \texttt{lock\_reorder}, the competitor will firstly check whether the lock is free (line 7).
If so, the competitor will enqueue immediately.
Otherwise, it will become a standby competitor.
During the reorder window, the standby competitors will occasionally check the lock's status (line 11).
We use a binary exponential back-off strategy (line 12) to reduce the contention over the lock
\footnote{
We present the non-blocking (busy-waiting) implementation here.
\design{} also has a blocking version that yields the thread using \texttt{nanosleep} during the reorder time window.
We evaluate both versions in Section~\ref{sec:eval}.
}.
When the reorder time window expires, the competitor can finally enqueue.
We do not use a secondary queue for the standby competitors because each competitor can have a different reorder window.
Thus, they may enqueue at a different time once the reorder window expires (not in FIFO order).
We also set an upper bound for the reorder window (omitted here) to make the reorderable lock starvation-free.
The reorder window is \textbf{not} a strict order constraint.
It is only a hint about the time the competitor can wait considering its latency SLO.
Thus, although big cores can still lock first when it expires (i.e., after invoking \texttt{lock\_fifo} in line 16 but have not enqueued yet),
it does not influence its correctness or efficiency.

Since the reorderable lock does not modify the underneath lock,
for unlocking, the reorderable lock directly invokes the unmodified unlock procedure \texttt{unlock\_fifo} (line 20).

\begin{lstlisting}[caption=Reorderable lock implementation., float, floatplacement=H, label=alg:reorder_impl]
int lock_immediately(mutex_t *mutex) {
  return lock_fifo(mutex);
}

int lock_reorder(mutex_t *mutex, u64 window) {
  u64 window_end, cnt = 0, next_check = 1;
  if (is_lock_free(mutex)) goto out;
  window_end = current() + window;
  while (current() < window_end) {
    if (cnt ++ == next_check) {
      if (is_lock_free(mutex)) goto out;
      next_check <<= 1;
    }
  }
out:
  return lock_fifo(mutex);
}

int unlock(mutex_t *mutex) {
  return unlock_fifo(mutex);
}
\end{lstlisting}

\subsection{\design{}}
\label{sec:libasl}

Atop of the reorderable lock,
\design{} collects the application's latency SLO and chooses a suitable reorder window accordingly to maximize the reordering without violating the SLO.
The mapping is achieved by tracing all epochs' latency and adjusting the reorder window at every epoch ends.
\design{} keeps individual reorder windows for each epoch.
When locking in an epoch,
\design{} calls \texttt{lock\_immediately} if running on big cores.
Otherwise, it calls \texttt{lock\_reorder} and sets the reorder window of that epoch.

\begin{lstlisting}[caption=\design{} epoch implementation., float, floatplacement=h, label=alg:epoch_impl]
typedef struct epoch {
  u64 window; /* Reorder Window*/
  u64 start;  /* Timestamp */
  u64 unit;   /* Adjust Unit */
} epoch_t;
__thread epoch_t epoch[MAX_EPOCH];
__thread int cur_epoch_id = -1;
__thread int *epoch_stack;
#define PCT 99 /* 99th Percentile Latency */

int epoch_start(int epoch_id) {
  if (cur_epoch_id >= 0)
    push(epoch_stack, cur_epoch_id);
  cur_epoch_id = epoch_id;
  epoch[epoch_id].start = current();
  return 0;
}

int epoch_end(int epoch_id, u64 SLO) {
  u64 latency, window;
  if (is_big_core()) goto out;
  latency = current()-epoch[epoch_id].start;
  window = epoch[epoch_id].window;
  if (latency > SLO) {
    window >>= 1;
    epoch[epoch_id].unit = window*(100-PCT)/100;
  } else {
    window += epoch[epoch_id].unit;
  }
  epoch[epoch_id].window = window;
out:
  cur_epoch_id = empty(epoch_stack) ? \
    -1 : pop(epoch_stack);
  return 0;
}
\end{lstlisting}

Algorithm~\ref{alg:epoch_impl} shows the implementation of the \design{}'s epoch interfaces.
Each epoch has the per-thread metadata,
which keeps the length of the reorder window (\texttt{window}), the start timestamp (\texttt{start}) and the adjusting unit of the epoch's length (\texttt{unit}).
When initializing, we give a default size to both the \texttt{window} and \texttt{unit}.
They will quickly adjust themselves to a suitable size after executing a few epochs.
When calling \texttt{epoch\_start}, \texttt{epoch\_id} specifies the unique id of the upcoming epoch,
which will be stored in the per-thread global variable \texttt{cur\_epoch\_id} (line 14).
To support the nested epoch, it pushes the outer epoch to the stack if it exists (line 13).
Then it records the start timestamp (line 15) by using the light-weight \texttt{clock\_gettime} ($\sim$45 cycles).

Besides the \texttt{epoch\_id}, \texttt{epoch\_end} takes another argument \texttt{SLO}, which specifies the latency SLO of the current epoch in nanoseconds.
It calculates the latency (line 22), compares it with the SLO (line 24) and updates the reorder window accordingly.
We take a conservative strategy to adjust the reorder window inspired by the TCP congestion control algorithm~\cite{allman1999tcp}, which combines linear growth and exponential reduction when latency exceeds.
We set the granularity of growth (\texttt{unit}) to be $\frac{100-PCT}{100}$
of the reduced window\footnote{
After another $\frac{100}{100-PCT}$ executions, the latency will be the same as the one which barely exceeds the SLO and triggers the exponential reduction. The probability of not exceeding the SLO is $(\frac{100}{100-PCT}-1)/\frac{100}{100-PCT} =                                                                               \frac{PCT}{100}$.
}, where \texttt{PCT} represents the percentile the SLO specifies (line 9, other percentiles are also supported).
It then checks the epoch stack to see whether there is a nested epoch and sets \texttt{cur\_epoch\_id} accordingly (line 32-33).

By leveraging weak-symbol replacement, 
\design{} redirects \texttt{pthread\_mutex\_lock} in applications
to \texttt{asl\_mutex\_lock} in Algorithm~\ref{alg:epoch_lock} transparently with negligible overhead (20+ cycles, similar to \texttt{litl}~\cite{litl}).
When calling \texttt{libASL\_lock}, competitors from big cores directly acquire the underneath lock using \texttt{lock\_immediately} (line 3),
while those from little cores use \texttt{lock\_reorder} and set the window length according to the current epoch (line 7-8).
If not in any epoch, the default maximum window is used to ensure that the thread will eventually lock (line 5).
Identifying the core type is achieved by getting the core id and looking up a pre-defined table.
Since the reorderable lock is implemented atop of existing locks, both the trylock and the nested locking are supported.
Besides, the conditional variable is also supported by using the same technique in \texttt{litl}~\cite{litl}.

\begin{lstlisting}[caption=\design{} internal interface., float, floatplacement=H, label=alg:epoch_lock]
int asl_mutex_lock(mutex_t *mutex) {
  if (is_big_core()) /* Big core */
    return lock_immediately(mutex);
  else if (cur_epoch_id < 0) /* Not in epoch */
    return lock_reorder(mutex, MAX_WINDOW);
  else /* In an epoch */
    return lock_reorder(mutex, 
      epoch[cur_epoch_id].window);
}
\end{lstlisting}

\subsection{Analysis}
\label{sec:libaslana}

\textbf{Throughput.}
\design{} provides good scalability in AMP.
We analyze different situations applications may encounter
and the corresponding behavior of \design{} as follows.

\emph{Big cores and little cores are \textbf{not} competing for the same lock.}
In big cores, \design{} behaves the same as the underneath FIFO lock (e.g., the MCS lock).
In little cores, \design{} behaves similarly to the backoff spinlock.
Both locks are scalable~\cite{boyd2012non} when competitors are from the same type of core.

\begin{figure*}[]
\centering
\subfloat[Performance comparison. \texttt{LibASL-X} means the SLO is set to \texttt{X}us. \texttt{LibASL-MAX} enables maximum reordering. \texttt{LibASL-OPT} uses static window.]
{
  \includegraphics[width=0.49\textwidth]{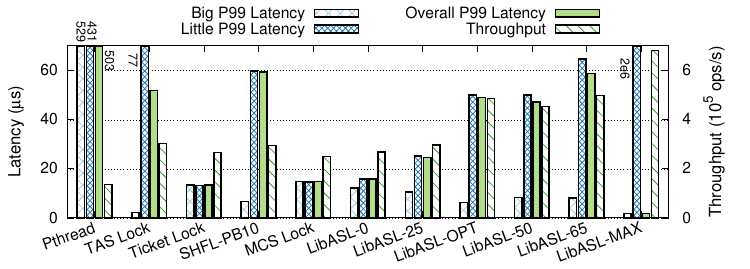}
  \label{fig:ch4_exp1}
}
\hspace{0.2em}
\subfloat[Variant SLOs (x-axis).]{
\includegraphics[width=0.233\textwidth]{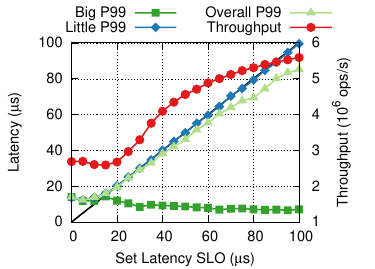}
\label{fig:ch4_exp2}
}
\subfloat[Mixing epochs with variant ratios. SLO is set to 100$\mu$s.]{
\includegraphics[width=0.233\textwidth]{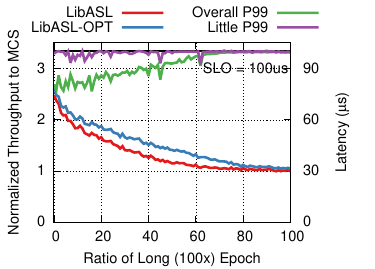}
\label{fig:ch4_ratio}
}

\hspace{0.2em}
\subfloat[Self-adaptive Reorder Window: Epochs' latencies during first 350ms.]{
\includegraphics[width=0.355\textwidth]{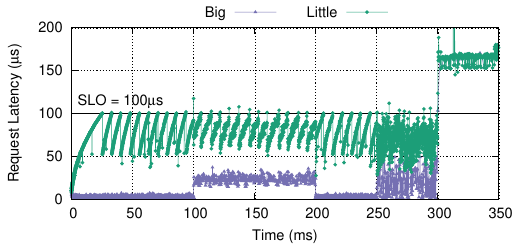}
\label{fig:ch4_exp4}
}
\hspace{0.2em}
\subfloat[Lock throughput.]{
\includegraphics[width=0.24\textwidth]{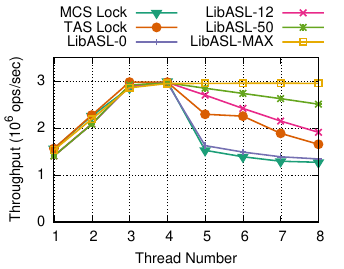}
\label{fig:ch4_scale}
}
\hspace{0.2em}
\subfloat[Overall tail latency from acquiring to releasing.]{
\includegraphics[width=0.24\textwidth]{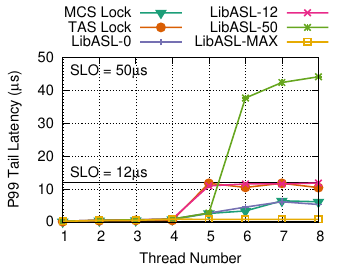}
\label{fig:ch4_scale_lat}
}

\vspace{-1em}
\hspace{0.2em}
\subfloat[Throughput speedup of \\ \design{} at variant contentions.]{
\includegraphics[width=0.204\textwidth]{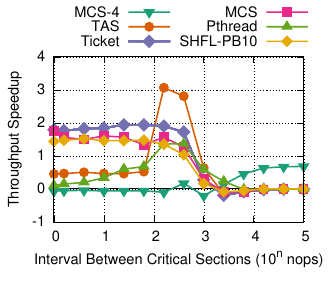}
\label{fig:ch4_exp3}
}
\hspace{0.1em}
\subfloat[Performance of blocking locks and \design{} when core-oversubscription. \texttt{LibASL-X} sets the SLO to \texttt{X}ms.]{
\includegraphics[width=0.4\textwidth]{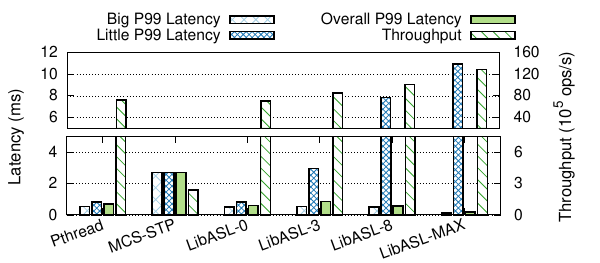}
\label{fig:ch4_exp5}
}
\hspace{0.1em}
\subfloat[\design{} with variant SLOs when core-oversubscription.]{
\includegraphics[width=0.233\textwidth]{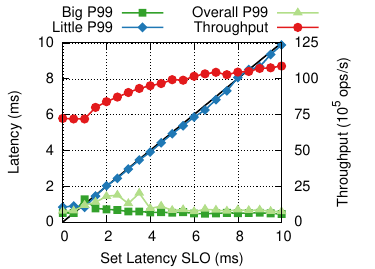}
\label{fig:ch4_exp6}
}
\hspace{0.2em}
\caption{Micro-benchmarks. \texttt{Big P99} and \texttt{Little P99} presents the 99th percentile latency in big and little cores individually.}
\vspace{-1em}
\label{fig:ch4_micro}
\end{figure*}

\emph{Big cores and little cores are competing for the same lock.}
When the lock is not heavily contented,
competitors from both big and little cores can immediately hold the lock if the lock is free (no additional overhead).
Little cores can help achieve higher throughput in such cases
(Figure~\ref{fig:ch4_exp3} shows the corresponding experiment).
With the contention level increased,
big cores will reorder with little cores under the condition that the latency SLO is still met.
When the big cores do not saturate the lock (i.e., the lock becomes free sometimes in a while),
little cores will lock once the queue is empty.
Thus, \design{} can find the sweet spot where some additional little cores help saturate the lock for better throughput (and block the rest little cores).
Otherwise, allowing any extra little core to join the competition will degrade the throughput.
In those cases, little cores can get the lock only when the reorder window expires.
Thus, \design{} can improve the throughput as much as the latency SLO allows.

\smallskip

\noindent\textbf{Latency.}
\design{} can precisely maintain the latency under SLO through a feedback mechanism.
The size of the reorder window has a monotonic relationship with the epoch's latency (i.e., a smaller window means a shorter waiting time).
It still holds when an epoch contains multiple lock acquisitions since they share the same window size.
Thus, \design{} can find the suitable window size that the latency barely meets the SLO by adjusting the size according to the latency (if the latency is higher than SLO, shrink the window, and vice versa).
Even if the epoch length (i.e., execution time) becomes heterogeneous (e.g., executing different code paths),
\design{} can still maintain the tail latency because the reorder window shrinks exponentially once the violation happens and grows linearly in the subsequent executions.
Thus, it only gives some short epoch a small reorder window (larger window can still meet SLO) but will not violate the SLO.

\design{} also supports threads migration or co-running with other applications.
Since each epoch's metadata is per thread,
they will not influence each other inherently.
Normally, the lock's contention remains stable in a while.
Thus, no extra window adjustment is required.
Otherwise,
when the thread gets scheduled out inside an epoch,
or the lock's contentions vary significantly, 
the window will quickly adjust itself in the same way as facing heterogeneous epochs.

Notice that the latency SLO is not a strict deadline.
\design{} uses it as a hint to maximize throughput without violating it.
There are three cases where \design{} does not take effect.
First, when the SLO is impossible to achieve even without reordering, \design{} falls back to a FIFO lock (best effort).
Second, when the SLO of nested epochs are mistakenly set (e.g., outer epoch has a tighter SLO), LibASL always prioritizes the inner epoch.
Third, if the workload is non-lock-sensitive (locks are barely contended), \design{} inherently will not influence its performance.

\smallskip
\noindent\textbf{Energy.}
Energy-efficiency is one of the major targets of the AMP system.
To maintain the lowest energy consumption,
Linux provides EAS (energy-aware scheduler~\cite{eassched}),
which chooses the most suitable core for each thread.
\design{} does not require core-binding, and threads can migrate between cores freely.
Therefore, \design{} will not violate the scheduling decision, as well as the energy target.
Threads will only run on big cores when the scheduler decides so (not caused by \design{}).
Moreover, when running on big cores,
\design{} makes threads do meaningful jobs rather than waiting,
which helps to save energy~\cite{energy}.

\smallskip
\noindent\textbf{Target systems.}
\design{} can be used in various AMPs.
Its improvement comes from considering the asymmetric computing capacity, and it does not restrict to a certain AMP.
Future AMPs may also have large core counts or even NUMA.
\design{} can adapt to those AMPs by replacing the underlying lock with the corresponding scalable locks (e.g., NUMA-aware locks).
It will prioritize big cores in the upper reorderable lock while achieving good scalability in the waiting queue of the underlying lock (e.g., NUMA-locality).

Some AMPs also support DVFS.
Although \design{} does not explicitly consider DVFS,
in most AMPs, a big core, even with the lowest OPP (Operating Performance Points), still has better performance than the highest little core OPP~\cite{yu2013power}.
Thus, \design{} still brings improvement by prioritizing faster big cores.
For other platforms (i.e., big cores may sometimes be slower than little cores), \design{} requires an extra mechanism to be aware of the performance variation due to DVFS.

\smallskip
\noindent\textbf{Limitations.}
\design{} brings improvement when the application is lock-sensitive and has a relatively loose SLO.
Otherwise, when the SLO is too tight (only achievable when passing the lock in FIFO),
or the workload is non-lock-sensitive, \design{} behaves the same as the underneath FIFO lock.

\design{} itself does not bring noticeable space or performance overhead.
The space overhead of \design{} comes from the metadata of epochs,
which is negligible because the per-thread metadata of an epoch only takes 24 bytes (see Algorithm~\ref{alg:epoch_impl}) and is irrelevant to the number of locks;
the two epoch operations only involve cheap computations ($\sim$93 cycles per epoch),
which barely influence the performance.

\section{Evaluation}
\label{sec:eval}

We evaluate \design{} to answer the following questions:

\begin{enumerate}
  \item How much throughput can \design{} improve when setting variant SLOs?
  \item Can \design{} precisely maintain epochs' latency in various situations?
  \item How does \design{} perform under variant contention levels?
  \item Can \design{} take effect in real-world applications?
\end{enumerate}

\noindent\textbf{Evaluation Setup.}
We evaluate \design{} on Apple M1, the only available desktop AMP yet.
M1 has 4 big cores and 4 little cores.
Similar to other AMP~\cite{biassched,pie},
their performances gap varies in different applications.
Big cores are 3.75x faster in Sysbench~\cite{sysbench},
while only 1.8x faster when executing the \texttt{NOP} instruction.
In such a system, the theoretical throughput speedup upper bound of \design{} to a FIFO lock (e.g., MCS) by prioritizing big cores is 1.8x\footnote{$\frac{4.75 + 1}{2}-1 \approx 1.8$: Comparing the case where the big cores always run and the case where the big cores and little cores run one by one.}.
We run Linux 5.11 on M1~\cite{m1-linux}.

Unless explicit statements, the reorderable lock is built atop the MCS lock, and the \texttt{PCT} is set to 99 (P99 latency).
We bind threads to different cores to evenly distribute them for stable results (not required by \design{}),
which is a widely-adopted evaluation method~\cite{mallock,CNA,flat,cohortlock,revisit,shuffle,cst,rcl,hclh,HBO,ffwd,sanl}.
We compare \design{} with pthread\_mutex\_lock (glibc-2.32), TAS, ticket, MCS and ShflLock~\cite{shuffle}.
ShflLock provides a lock reordering framework but can only take a static policy (e.g., NUMA-local policy prioritizes competitors from the same node).  
The SLO-guided ordering in \design{} is not static.
Thus it is hard to be integrated.
Instead, we adopt a proportional-based static policy, which gives the big core a fixed higher (10x) chance to lock.
It is implemented by modifying the existing NUMA-local policy to separate asymmetric cores into two nodes.
Rather than passing in each node evenly (for preserving long-term fairness in NUMA-local policies), it uses a simple counter to allow exactly 1 little core to lock after every 10 big cores.
Since any proportion is one static trade-off between latency and throughput (Figure~\ref{fig:ch2_proportion}),
we choose the proportion (i.e., 10) that has obvious throughput improvement without introducing extremely long latency.
We also present the speedup upper bound of \design{} by using the maximum reorder window (100ms).

\subsection{Micro-Benchmarks}
\label{sec:microbench}


\textbf{Bench-1:} \emph{A heavily contended benchmark.} 
In this benchmark, all threads repeatedly execute the same epoch,
which contains 4 critical sections of different lengths protected by 2 different locks.
In each critical section, threads read-modify-write a specific number of shared cache lines (64 in total).
We insert a fixed number (600*$2^7$) of \texttt{NOP} instructions between two epochs.
We present epoch's tail latency in little cores, big cores and overall separately.

As shown in Figure \ref{fig:ch4_exp1},
the TAS lock shows big-core-affinity here (big cores have a shorter tail latency).
Thus, it has the highest throughput yet the longest latency among existing locks.
When setting the SLO to 0 (\texttt{LibASL-0}),
\design{} performs the same as the MCS lock since the SLO is impossible to achieve (falls back to FIFO).
Compared with the TAS lock,
when achieving a similar throughput (\texttt{LibASL-25}),
\design{} reduces the tail latency by over 50\%;
when having a similar tail latency (\texttt{LibASL-50}),
\design{} achieves 50\% better throughput.
Although TAS lock also implicitly prioritizes the big cores here,
the reordering depends on the hardware and is unstable and uncontrollable.
\design{} manages the reordering elaborately,
which allows more critical sections to execute on the big cores and thus has higher throughput. 
\design{} brings up to 1.2x speedup to TAS lock (1.7x to MCS) when setting a larger SLO (\texttt{LibASL-MAX}).
Although the proportional-based approach (\texttt{SHFL-PB10}) performs better than MCS by 20\%, it has a 4x longer tail latency.
\design{} outperforms it by 70\% when having similar tail latency (\texttt{LibASL-65}).
It is because \design{} has a better cache locality by batching more big cores before passing to little cores for no SLO violation.
In contrast, the proportional-based approach has to periodically (i.e., every 10 big cores) hand over the lock to little cores.
The pthread\_mutex\_lock has the worst throughput and the longest latency.
\design{} outperforms it by 4x at most.
(\textbf{Question 1})

We also compare \design{} with the optimal policy \texttt{LibASL-OPT}, which directly chooses a static window (no window adjustment) to present \design{}'s performance headroom.
When having a similar tail latency (\texttt{LibASL-50}),
the cost of the window adjustment is only 6\%.

Figure~\ref{fig:ch4_exp2} shows \design{}'s performance in \textit{Bench-1} when setting variant SLOs.
As shown in the figure, with setting a larger SLO (x-axis from left to right),
the throughput increases, and the tail latency of little cores sticks straightly to the Y=X line (i.e., barely meet the SLO).
Meanwhile, since big cores get more chances to lock,
their latencies are much shorter.
The growth speed of throughput slows down with the SLO becoming larger.
It is intuitive because the benefit of reordering more big cores will decrease if most critical sections are already executed on the faster big cores.
The only exception is when setting an SLO shorter than 15$\mu$s (the tail latency of MCS in Figure~\ref{fig:ch4_exp1}),
the SLO is impossible to achieve.
Therefore, \design{} falls back to the MCS lock.

\smallskip
\noindent\textbf{Bench-2:} \emph{A highly variable workload.}
We record each epoch's latency in the first 350ms when executing \emph{Bench-1}.
Figure \ref{fig:ch4_exp4} shows the latency of each epoch executed on big and little cores individually.
The SLO is set to 100$\mu$s.
During the 100 and 200ms period, we enlarge the epoch's length by 128 times (by accessing more cache lines) and shrink it back to the original length during the 200 and 250ms period.
After that, we change the length of each epoch randomly (access a random number of cache lines) during the 250 to 300ms period.
Finally, the epoch's length is set 1024 times longer till the end.
As shown in the figure, \texttt{LibASL} is fully capable of maintaining the latency in a highly variable workload (\textbf{Question 2}).
Every time the latency exceeds the SLO, the reorder window shrinks to its half and increases gradually.
When the epoch's length changes at 100 ms and 200 ms,
\design{} quickly adjusts the reorder window to a suitable size.
Even when the length becomes highly heterogeneous during 250 and 300 ms,
\design{} can still keep the latency within the SLO.
When the epoch's length enlarges 1024 times at 300 ms,
the SLO is impossible to achieve.
Thus, \design{} falls back to FIFO,
and both big and little cores have similar latencies.

\smallskip
\noindent\textbf{Bench-3:} \emph{A benchmark mixing epoch of significantly different lengths.}
In Figure~\ref{fig:ch4_ratio}, we randomly generate short and long (100$\times$ longer by inserting more \texttt{NOP} instructions) epochs of different ratios (e.g., x=20 means 20\% of epochs are short while 80\% are long).
We compare \design{} with \texttt{LibASL-OPT}, which directly chooses a suitable (static) window for different epochs (impossible in the real world).
The SLO is set to 100$\mu$s throughout the experiment.
As shown in the figure,
\design{} brings significant and close-to-optimal (maximum 20\% gap with \texttt{LibASL-OPT} at 50\% ratio) throughput improvement to MCS while precisely keeping the latency within SLO in all ratios (\textbf{Question 2}).
When all epochs are long (i.e., x=100),
the tail latency of the MCS lock is also 100$\mu$s (the same as the SLO).
Thus, \design{} falls back to FIFO for not violating SLO and has the same throughput (i.e., y=1) as the MCS lock.

\smallskip
\noindent\textbf{Bench-4:} \emph{Scalability.}
We present the scalability of \design{} using the same benchmark setup as Figure~\ref{fig:ch2_moti}.
As shown in Figure~\ref{fig:ch4_scale} and \ref{fig:ch4_scale_lat},
when setting the SLO to 0,
\texttt{LibASL-0} behaves the same as the MCS lock.
When setting the SLO to have the same tail latency (12$\mu$s) as the TAS lock,
\texttt{LibASL-12} achieves better throughput scalability.
The throughput of \texttt{LibASL-MAX} does not drop at all.
Due to the high contention, it barely passes the lock to the little cores.

\begin{table}[]
\centering
\caption{Databases Considered}
\vspace{-1em}
\resizebox{1\columnwidth}{!}{
\begin{tabular}{lllll}
\toprule
Application    & Benchmark  & Locks in each Epoch   \\ \midrule
\begin{tabular}[c]{@{}l@{}}\textbf{Kyoto Cabinet}~\cite{kcdb}\\In-memory KV\end{tabular} & 50\% Put 50\% Get &
\begin{tabular}[c]{@{}l@{}}Slot-level Lock\\Method Lock\end{tabular}\\
\begin{tabular}[c]{@{}l@{}}\textbf{upscaledb}~\cite{upscaledb}\\On-disk KV\end{tabular}  & 50\% Put 50\% Get & 
\begin{tabular}[c]{@{}l@{}}Global Lock\\Worker Pool Lock\end{tabular}\\
\begin{tabular}[c]{@{}l@{}}\textbf{LMDB}~\cite{chu2011mdb,LMDB}\\On-disk KV\end{tabular} & 50\% Put 50\% Get & 
\begin{tabular}[c]{@{}l@{}}Global Lock\\Metadata Locks\end{tabular}\\
\begin{tabular}[c]{@{}l@{}}\textbf{LevelDB}~\cite{leveldb}\\On-disk KV\end{tabular}	& \texttt{db\_bench} Random Read & Metadata Lock\\
\begin{tabular}[c]{@{}l@{}}\textbf{SQLite}~\cite{SQLite}\\On-disk Database\end{tabular}
      & \begin{tabular}[c]{@{}l@{}}1/3 Insert 1/3 Simple Select\\ 1/3 Complex Select\end{tabular}
 &  \begin{tabular}[c]{@{}l@{}}State Machine Lock\\Metadata Locks\end{tabular}
 \\
\toprule           
\end{tabular}
\vspace{-1em}
}
\label{tlb:application}
\end{table}

\smallskip
\noindent\textbf{Bench-5:} \emph{A benchmark with variant contentions.}
In this benchmark, threads acquire the same lock to read-modify-write 2 shared cache lines.
We alter the contention by executing a different number of \texttt{NOP} instructions between two lock acquisitions.
Figure \ref{fig:ch4_exp3} shows the throughput speedup of \design{} over the locks in the legends 
(e.g., when x=0, \design{} outperforms MCS by 2x and TAS lock by 45\%).
To allow the maximum reordering, we do not set SLO in \design{}.
We also include the result of only using big cores (only \texttt{MCS-4}, other locks use all the cores).
When competitors from big cores already saturate the lock (x $<$ 3),
\design{} makes the competitors standby and 
achieves similar throughput with \texttt{MCS-4} (significantly better than others).
With the contention level decreased, \design{} allows little cores to join the competition and achieves 68\% better throughput than only using big cores.
It also validates that little cores can bring noticeable improvement in some loads.
Among all contention levels, \design{} achieves good throughput.
(\textbf{Question 3})

\smallskip
\noindent\textbf{Bench-6:} \emph{A benchmark with CPU core over-subscription}.
We examine the effectiveness of \design{} in a core over-subscription situation by creating 2 threads on each core and executing \emph{Bench-1}.
We replace the non-blocking MCS lock in \design{} with the pthread\_mutex\_lock and use \texttt{nanosleep} to replace the busy waiting of the standby competitor (i.e., line 9 in Algorithm~\ref{alg:reorder_impl}, the sleep time is set in a back-off manner).
Results are presented in Figure \ref{fig:ch4_exp5} and \ref{fig:ch4_exp6}.
Since the MCS lock passes the lock in a FIFO order,
the waking-up latency will be introduced to the critical path, leading to a significant throughput degradation (spin-then-park MCS is 96\% worse than pthread\_mutex\_lock).
Therefore, \design{} uses pthread\_mutex\_lock rather than the spin-then-park MCS lock.
Although pthread\_mutex\_lock does not guarantee the FIFO order and thus has an unstable lock acquisition latency,
\design{} still can preserve the SLO owing to its self-adaptive reorder window
and outperform the pthread\_mutex\_lock by up to 80\%.

\subsection{Application Benchmarks}

\label{sec:appbench}

We evaluate 5 popular databases in Table \ref{tlb:application} to show the effectiveness of \design{} in the real-world (\textbf{Question 4}).
Databases benefit from using little cores to handle more requests in fewer machines,
which can improve the cost and energy efficiency in edge computing~\cite{WorkloadCompactor,dvfs,AndrewCake}.
Integrating \design{} only requires inserting 3 lines of code: annotating the epoch with \texttt{epoch\_start} and \texttt{epoch\_end} and adding the header file.
As prior work~\cite{CNA,shuffle} does,
we run each benchmark for a fixed period and calculate the average throughput.
Moreover, to present the effectiveness of \design{} in cases where epochs' lengths are highly heterogenous,
we randomly choose to insert or find 1 item (fifty-fifty, referring to YCSB-A~\cite{ycsb-workload}) in an epoch.
In most benchmarks, each epoch acquires multiple locks as listed in the rightmost column of Table \ref{tlb:application}.
For each database,
we first set several specific SLOs to present a performance comparison with existing locks 
to show the improvement of \design{}
when having similar latency or throughput with existing locks (e.g., TAS).
Then we show the performance under other SLOs in \textit{Variant SLOs} figures.

\begin{figure}[]
\centering
\subfloat[Kyoto Cabinet. 1e6 means $10^6$. The chosen SLOs are only for easier comparing. Other settings are detailed in figure~\ref{fig:ch4_kc_exp2}.]{
  \includegraphics[width=0.5\textwidth]{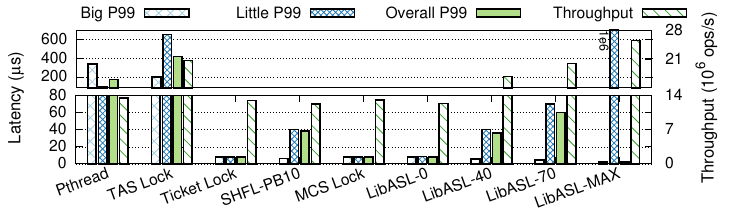}
  \label{fig:ch4_kc_exp1}
}

\vspace{-0.5em}
\subfloat[Variant SLOs]{
\includegraphics[width=0.236\textwidth]{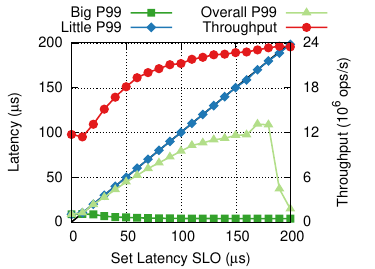}
\label{fig:ch4_kc_exp2}
}
\subfloat[CDF (SLO: 70$\mu$s)]{
\includegraphics[width=0.1824\textwidth]{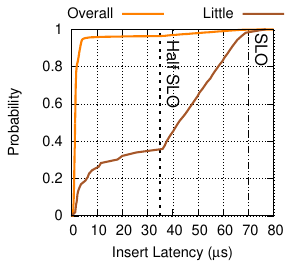}
\label{fig:ch4_kc_cdf}
}

\vspace{-0.5em}
\subfloat[upscaledb]{
  \includegraphics[width=0.5\textwidth]{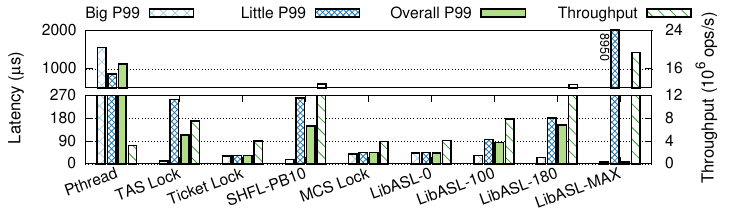}  \label{fig:ch4_ups_exp1}
}

\vspace{-0.5em}
\subfloat[Variant SLOs]{
  \includegraphics[width=0.236\textwidth]{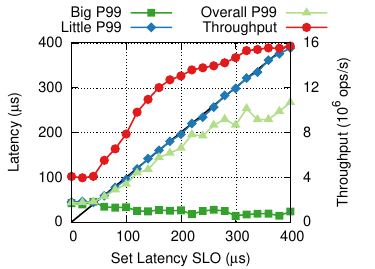}
  \label{fig:ch4_ups_exp2}
}
\subfloat[CDF (SLO: 140$\mu$s)]{
\includegraphics[width=0.1824\textwidth]{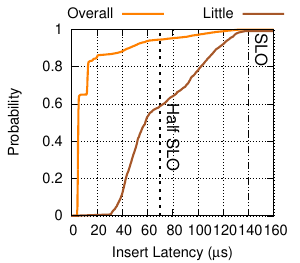}
\label{fig:ch4_ups_cdf}
}

\vspace{-0.5em}
\subfloat[LMDB. 5e5 means $5\times10^5$.]{
  \includegraphics[width=0.5\textwidth]{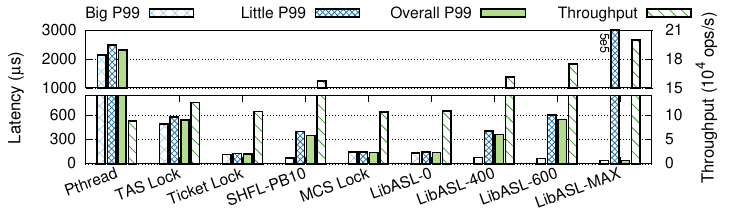}
  \label{fig:ch4_lmdb_exp1}
}

\vspace{-0.5em}
\subfloat[Variant SLOs]{
  \includegraphics[width=0.236\textwidth]{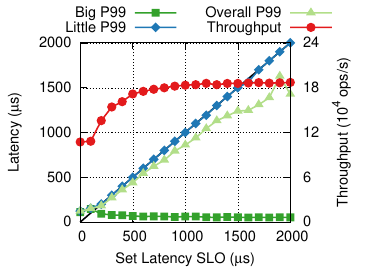}
  \label{fig:ch4_lmdb_exp2}
}
\subfloat[CDF (SLO: 1900$\mu$s)]{
\includegraphics[width=0.1824\textwidth]{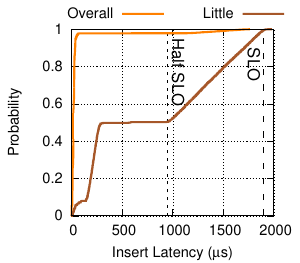}
\label{fig:ch4_lmdb_cdf}
}
\vspace{-0.5em}
\caption{
Databases.
Legends are explained in Figure~\ref{fig:ch4_micro}.
}
\end{figure}

\begin{figure*}[]
\centering
\subfloat[LevelDB. 6e5 means $6\times10^5$.]{
  \includegraphics[width=0.55\textwidth]{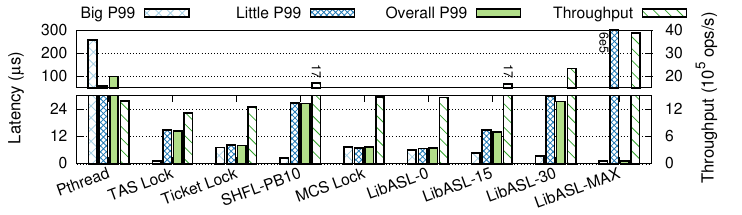}
  \label{fig:ch4_ldb_exp1}
}
\hspace{-1em}
\subfloat[Variant SLOs]{
  \includegraphics[width=0.25\textwidth]{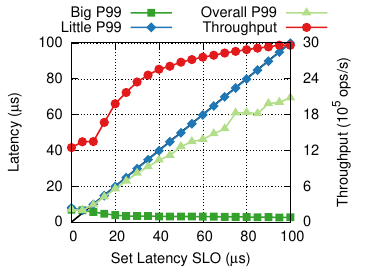}
  \label{fig:ch4_ldb_exp2}
}
\hspace{-1em}
\subfloat[CDF (SLO: 100$\mu$s)]{
\includegraphics[width=0.2\textwidth]{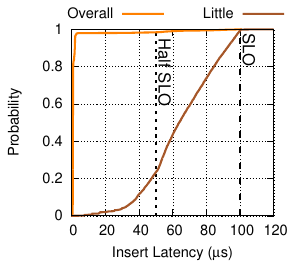}
\label{fig:ch4_ldb_cdf}
}

\vspace{-1em}
\subfloat[SQLite. \texttt{LibASL-X} means the SLO is set to \texttt{X} ms.]{
  \includegraphics[width=0.55\textwidth]{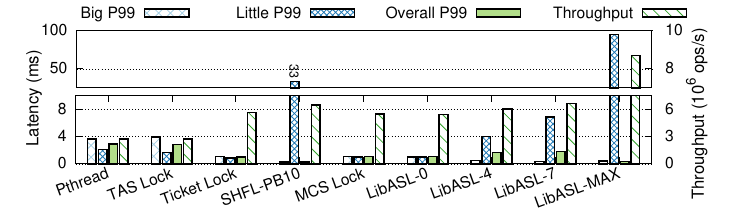}
  \label{fig:ch4_sqlite_exp1}
}
\hspace{-1em}
\subfloat[Variant SLOs]{
  \includegraphics[width=0.25\textwidth]{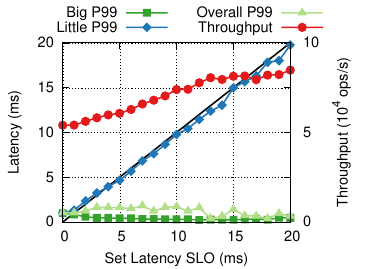}
  \label{fig:ch4_sqlite_exp2}
}
\hspace{-1em}
\subfloat[CDF (SLO: 4ms)]{
\includegraphics[width=0.2\textwidth]{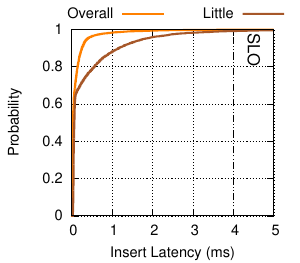}
\label{fig:ch4_sqlite_cdf}
}
\vspace{-1em}
\caption{
Database benchmarks.
Legends are explained in Figure~\ref{fig:ch4_micro}.
}
\vspace{-1em}
\end{figure*}

We first evaluate \design{} in several KV-stores.
KV-store plays an important role in CDN or IoT edge servers as the storage service~\cite{Akamai, IoTDB}.
As shown in Figure \ref{fig:ch4_kc_exp1},
in Kyoto Cabinet,
\design{} reduces the tail latency by 90\% when achieving similar throughput with the TAS lock (\texttt{LibASL-70})
and improves the throughput by up to 23\% with a larger SLO (96\% to MCS, 89\% to pthread\_mutex\_lock).
When having a similar latency with \texttt{SHFL-PB10} (\texttt{LibASL-40}),
\design{} improves the throughput by 38\%.
Figure \ref{fig:ch4_kc_exp2} shows the performance of \design{} when setting variant SLOs.
Although the execution time of \texttt{Put} or \texttt{Get} is heterogeneous,
\design{} can still precisely maintain the tail latency.

Figure \ref{fig:ch4_kc_cdf} presents the latency Cumulative Distribution Function (CDF) of \design{} when setting the SLO to 70$\mu$s.
\texttt{Overall} and \texttt{Little} represent the overall and little core's latency.
A clear boundary can be seen in the overall latency since most operations are executed on big cores.
Due to the intensive contention,
only less than 20\% of operations are executed on little cores and have longer latency.
There is also a clear boundary in little core's latency.
About half of the operations have shorter latency (<35$\mu$s) due to the shorter execution time of \texttt{Get}.
As for the longer \texttt{Put} operation (another half),
since the reorder window shrinks by half and grows linearly once the latency exceeds the SLO,
the probability also grows linearly after the \textit{Half SLO} (35$\mu$s).


Upscaledb and LMDB show similar results
(Figure \ref{fig:ch4_ups_exp1}$\sim$\ref{fig:ch4_lmdb_cdf}).
In upscaledb, the TAS lock shows big-core affinity.
Thus it has 90\% higher throughput yet 2.5x longer tail latency than the MCS lock.
When having a similar tail latency (\texttt{LibASL-140}),
\design{} improves the throughput of the TAS lock by 46\%,
which further goes to 1.6x (3.8x to MCS, 50\% to \texttt{SHFL-PB10} and 5x to pthread\_mutex\_lock) with a larger SLO.
In LMDB,
\design{} outperforms the TAS lock by 40\% when having a similar latency (\texttt{LibASL-600}), which goes to 60\% (86\% to MCS, 27\% to \texttt{SHFL-PB10} and 126\% to pthread\_mutex\_lock) with a larger SLO.
In the latency CDFs of both benchmarks (Figure~\ref{fig:ch4_ups_cdf}, \ref{fig:ch4_lmdb_cdf}),
the clear boundary in little core's latency distinguishes the shorter \texttt{Get} from the longer \texttt{Put} operation.

LevelDB is another widely-used KV-store.
However, LevelDB implements its own blocking strategy rather than directly using the pthread\_mutex\_lock on \texttt{Put} operation.
Thus we use the \texttt{randomread} test in the build-in \texttt{db\_bench} to only test its \texttt{Get} operation,
which will acquire a global lock to take a snapshot of internal database structures.
As shown in Figure \ref{fig:ch4_ldb_exp1},
\design{} improves the throughput of TAS lock by 50\% when having a similar latency (\texttt{LibASL-15}), which goes to 2.5x (1.6x to MCS, 1.8x to pthread\_mutex\_lock and 1.3x to \texttt{SHFL-PB10}) with a larger SLO.
Since we only test the \texttt{Get} operation, 
most requests have a longer latency than the \textit{Half SLO} as shown in Figure \ref{fig:ch4_ldb_cdf}.

Finally, we evaluate \design{} in SQLite, which has been used in Azure IoT edge servers~\cite{Azure}.
We place 1/3 \texttt{Insert}, 1/3 simple query (point query on an indexed column) and 1/3 complex query (range query on an indexed column with a filter on a non-indexed column) in a \texttt{DEFERRED} transaction enclosed in an epoch.
Moreover, we add an extremely long full-table scan on a 100k table every 1000 executions in the same epoch to present that \design{} can survive on some occasionally appeared extremely long requests.
SQLite uses locks to protect the internal state machine.
The transaction can commit successfully only in a certain state.
Thus epochs' latencies greatly fluctuate and grow non-linearly in Figure~\ref{fig:ch4_sqlite_cdf}.
It also widens the gap of the transaction's success rate in big and little cores when using \texttt{SHFL-PB10} and causes a latency collapse in little cores.
Both the simple and the complex \texttt{Select} operations have a much shorter execution time than the \texttt{Insert} operation.
Thus, 2/3 of the requests have shorter tail latency (latency grows significantly after y$>$2/3 in Figure~\ref{fig:ch4_sqlite_cdf}).
However, even with some occasionally appeared extremely long epochs, \design{} still can precisely keep the tail latency under the SLO and improve the throughput as shown in Figure \ref{fig:ch4_sqlite_exp2}.
\design{} brings up to 2.1x speedup to the TAS lock (55\% to MCS, 2.1x to pthread\_mutex\_lock and 35\% to \texttt{SHFL-PB10}) without violating SLO.

Besides M1, we also evaluated \design{} in Hikey970~\cite{hikey970} (ARM big.LITTLE) and a simulated Intel AMP (through per-core DVFS).
\design{} works well on both platforms since the improvement comes from considering the asymmetry in computing capacity and is \textbf{not} restricted to a certain AMP.
Specifically, \design{} brings 34$\sim$94\% (Intel) and 37$\sim$87\% (Hikey970) throughput improvement to the MCS lock while preciously maintaining the SLO in the same database benchmarks.
Detailed results are omitted due to the space limit.

\subsection{Evaluation Highlights}

Results confirm the effectiveness of \design{} in AMP.
First, \design{} can precisely maintain the latency SLO even in highly variant workloads.
Second, \design{} shows promising performance advantages over existing locks.
Compared to fair locks (lowest latency but low throughput),
\design{} can significantly improve the throughput (e.g., 3.8x to the MCS lock in upscaledb).
Compared to unfair locks (highest latency but sometimes high throughput),
\design{} has much lower tail latency when achieving similar throughput 
(e.g., 90\% lower than the TAS lock in KyotoCabinet),
and substantially higher throughput when ensuring similar tail latency
(e.g., 46\% higher than the TAS lock in LMDB).
Moreover, \design{} can further outperform the TAS lock when setting a larger SLO (e.g., by 2.5x in LevelDB).
Compared to the static proportional-based approach,
\design{} can better meet applications' needs by improving the throughput as much as possible considering a certain latency SLO that a fixed proportion cannot.

\section{Related Work}
\label{sec:diss}

Scalable synchronization primitives~\cite{CNA,shuffle,cohortlock,mallock,flat,rcl,ffwd,sanl,flatmcs,hclh,revisit,psim,oyama1999executing,ahmcs,hmcs,HBO,self-tuningTAS, mutable, spin-then-parklock, prwlock, prwlock-tpds, Pisces, CLoF, vsync, pilot}
have been extensively studied over decades, targeting various scenarios.
Yet, there lacks investigation on the scalability problem on AMP.

There are already some locks~\cite{CNA,shuffle,cohortlock,mallock,flatmcs,hclh,ahmcs,HBO, CLoF} that reorder competitors to achieve better throughput in NUMA or many-core processors, which inspire \design{}.
Among those locks, defining an effective lock ordering (i.e., how to reorder) to achieve different optimization targets is important and challenging.
The major difference between \design{} and those locks is that \design{} defines a new SLO-guided AMP-specific ordering while existing lock orderings (e.g., NUMA-local) are non-scalable in AMP.
As analyzed in Section~\ref{sec:analyze}, those locks preserve the long-term fairness to keep a relatively low latency, which brings throughput collapse on AMP.
Besides, unlike \design{}, the long-term fairness only forbids starvation, leaving the latency unpredictable.
Rather than proposing a specific ordering,
ShflLock~\cite{shuffle} provides a reordering framework.
It relies on a provided \emph{static} policy to shuffle the queue internally.
However, the scalability issue in AMP cannot be easily solved by proposing a static policy.
Preserving fairness brings throughput collapse;
reordering without limit causes latency collapse;
proportional execution brings unstable and unpredictable performance so that a suitable static proportion is hard to find as evaluated in Section~\ref{sec:eval}.
Instead, \design{} has a \emph{dynamic} priority to achieve good AMP scalability.

Delegation locks~\cite{flat,rcl,ffwd,sanl,revisit,psim,oyama1999executing} reduce the data movement by executing all critical sections in one core (the lock server), which significantly improves the throughput in NUMA.
Although placing the lock server on big cores can hide the weak computing capacity of little cores,
it requires the big core busy polling,
which wastes a precious big core and violates the energy target at lower contention.
A more severe obstacle to using the delegation lock is that it requires non-trivial code modifications to convert all critical sections into closures,
which brings enormous engineering work due to the complexity of real-world applications.
Instead, \design{} only requires linking and, if latency-critical, inserting few lines of code to specify the latency SLO.

There are also self-tuning locks~\cite{self-tuningTAS, mutable, spin-then-parklock} that tune some internal parameters at runtime.
\design{} can be regarded as one such lock considering its self-adaptive reorder window.
However, existing self-tuning locks target solving different problems and thus have different designs.
Specifically, reactive lock~\cite{self-tuningTAS} tunes the back-off time in spinlock to reduce the contention;
mutlock~\cite{mutable} tunes the number of busy spinning threads to hide the wakeup latency in blocking locks (also named as \textit{window}, but it is a different mechanism:
its \textit{spinning window} is per-lock and restricts how many threads can busy spin;
\design{}'s \textit{reorder window} is per-thread and controls how long others can reorder);
locks in \cite{spin-then-parklock} tune the spinning threshold before parking for better throughput.
Unlike them,
\design{} tunes the reorder window to preserve a new AMP-aware lock ordering for better scalability on AMP.
Existing self-tuning locks do not preserve a specific lock ordering and fail to scale on AMP.

\design{} blocks some competitors from little cores to prioritize big cores.
Some existing locks~\cite{mutable,spin-then-parklock,mallock} also leverage concurrency restrictions for different targets.
Mutlock~\cite{spin-then-parklock} and locks in \cite{spin-then-parklock} park some competitors to save CPU;
Malthusian lock~\cite{mallock} only allows a subset of competitors to lock for reducing contention.
Those locks target SMP, and their techniques cannot solve the scalability issue in AMP. 
Facing the asymmetry,
it is challenging to restrict certain competitors for better performance.
\design{} adopts an SLO-guided runtime restriction (through reorder window) and achieves good scalability in AMP.

Computing capacity can also be asymmetric in SMP when using DVFS.
Previous work~\cite{TURBO, dvfs-lock-acc, dvfs-identify-cs} boosts the frequency of the lock holder to gain better throughput.
However, unlike DVFS, the asymmetry in AMP is inherent.
Thus, those techniques also cannot be applied to AMP.

Improving the system's throughput for better cost and energy efficiency without violating the latency SLO is a widely adopted technique~\cite{WorkloadCompactor,AndrewCake,Elfen-Scheduling}.
WorkloadCompactor~\cite{WorkloadCompactor} and Cake~\cite{AndrewCake} reduce the datacenter's cost by consolidating more loads into fewer servers without violating the latency SLO.
Elfen Scheduling~\cite{Elfen-Scheduling} leverages the SMT to run latency-critical and other requests simultaneously to improve the utilization without compromising the SLO.
\design{} takes a similar approach to solve the lock scalability issue in AMP.

\section{Conclusion}

In this paper, we propose an asymmetry-aware scalable lock named \design{}.
It provides a new latency SLO-guided lock ordering to prioritize big cores for better throughput while elaborately maintaining little cores' latencies.
Evaluations on real-world applications show that \design{} brings better performance over counterparts on AMP.

\section{Acknowledgement}

We sincerely thank all the anonymous reviewers for their insightful suggestions.
This work is supported in part by China National Natural Science Foundation (No. 61925206), 
High-Tech Support Program from Shanghai Committee of Science and Technology (No. 19511121100), and Huawei.
Jinyu Gu is the corresponding author.

\bibliographystyle{ACM-Reference-Format}
\bibliography{arxiv-libasl}



\end{document}